# On Digital Subcarrier Multiplexing under A Bandwidth Limitation and ASE Noise

JUNHO CHO,* XI CHEN, GREG RAYBON, AND SON THAI LE

*Nokia Bell Labs, 600 Mountain Ave, Murray Hill, NJ 07974, USA*
*[junho.cho@nokia-bell-labs.com](mailto:junho.cho@nokia-bell-labs.com)*

**Abstract:** We show that digital subcarrier multiplexing (DSM) systems require much greater complexity for Nyquist pulse shaping than single-carrier (SC) systems, and it is a misconception that both systems use the same bandwidth when using the same pulse shaping. Through back-to-back (B2B) experiments with realistic transmitter (TX) modules and amplified spontaneous emission (ASE) noise loading, we show that even with optimized water-filling and entropy loading, DSM does not achieve a larger net data rate (NDR) compared to SC when only ASE noise exists in the channel in long-haul transmission scenarios.

## 1. Introduction

Digital subcarrier multiplexing (DSM) [1-6] can in principle increase the achievable net data rate (NDR) of coherent optical transmission systems. It is because: *(i)* it can reduce nonlinear interference in optical fiber by digitally realizing a near-optimal symbol rate (per subcarrier), which in certain cases may be lower than the symbol rate of single-carrier (SC) transceivers [1-4], *(ii)* it allows to adapt the data rate on a per-subcarrier basis using a hybrid modulation format [3] or entropy loading [5,6], and *(iii)* it can reduce the performance degradation due to equalization-enhanced phase noise (EEPN) in long-haul transmission when laser linewidth is excessively large [7-9]. Through simulations and experiments, several studies have reported the performance improvement achieved by DSM over SC [1,3,5]. However, further studies are needed on the effect of bandwidth limitations of real-world optoelectronic modules on the performance of DSM and the implementation complexity of transmitter (TX) and receiver (RX) digital signal processing (DSP) for DSM.

In this work, we show that DSM requires much greater complexity for Nyquist pulse shaping in TX DSP to achieve similar performance as SC. We show by experiments that even allowing greater TX DSP complexity for DSM, DSM provides little or no increase in achievable NDR compared to SC, when only amplified spontaneous emission (ASE) noise is present in the channel. We perform ASE noise-loaded back-to-back (B2B) experiments using realistic TX modules that jointly lead to gradually decreasing gain towards high frequencies. In theory, DSM with water-filling [10,11] and entropy loading [5] has been expected to result in NDR improvement in this condition. In practice, however, our experimental results show that even with optimized water-filling and ideal entropy loading, DSM does *not* achieve a distinctively larger NDR compared to SC. As reasons for the discrepancy between the theory and practice, we show that: *(i)* it is a misconception that DSM uses the same bandwidth as SC when both systems use the same Nyquist pulse shaping, but the overall signal bandwidth is expanded by DSM due to the spectral margins between subcarriers, *(ii)* pre-emphasis filtering is essential to maximize NDR, but it makes TX noise vary with the filter shape instead of being added linearly with a fixed power, thus invalidating the assumption for optimality of water filling and entropy loading, and *(iii)* DSM yields only insignificantly greater NDRs than SC in the absence of ASE noise, and this gain disappears as the inline ASE noise increases.

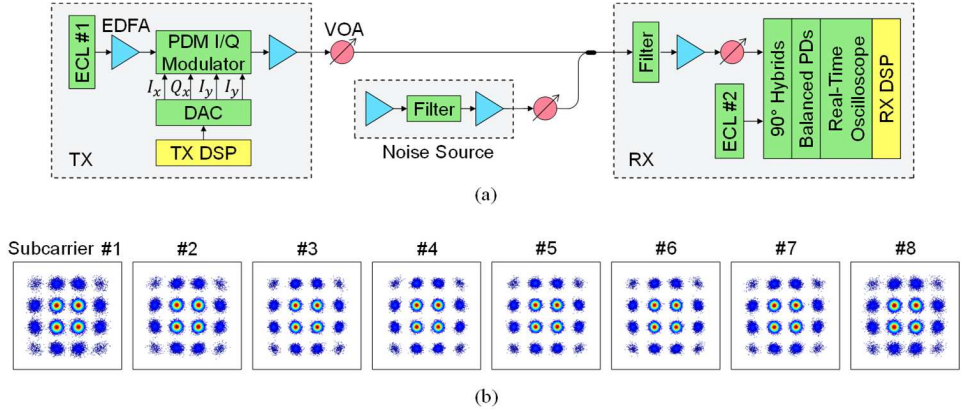

Fig. 1. (a) Experimental setup, and (b) typical recovered constellations at a high OSNR.

## 2. Experiments

### 2.1 Experimental setup

Fig. 1 shows the experimental system used in this work. At the TX, a semiconductor external cavity laser (ECL) with linewidth <100 kHz generates a carrier at 193.4 THz. A 4-channel complementary metal-oxide-semiconductor (CMOS) digital-to-analog converter (DAC) with a 6-dB bandwidth of ~35 GHz at 120 GSa/s feeds signals to a polarization-division multiplexing (PDM) in-phase and quadrature (I/Q) Mach-Zehnder modulator (MZM). ASE noise is added to the signal by two cascaded erbium-doped fipber amplifiers (EDFAs). An optical filter between them and another optical filter at the RX limit the noise bandwidth to 500 GHz around 193.4 THz. Variable optical attenuators (VOAs) realize a target optical signal-to-noise ratio (OSNR) in an automated manner by controlling the signal and noise powers. In the RX, the same type of ECL as in the TX is used as a local oscillator (LO). Detection is performed using four balanced photodiodes (BPDs) with a 3-dB bandwidth of 75 GHz, followed by a 4-channel real-time oscilloscope operating at 256 GSa/s with an analog bandwidth of 103 GHz. Since the RX has a much wider bandwidth than the TX and a nearly frequency-flat response, the bandwidth limitation is imposed by the TX in this setup.

### 2.2 DSP for SC and DSM systems

We use the typical TX and RX DSP chains shown in Fig. 2(a) for SC and Fig. 2(b) for DSM. We transmit probabilistically shaped (PS) 16-ary quadrature amplitude modulation (QAM) symbols at a shaping rate of 3.2 b/sym/pol.

In the TX DSP for SC (Fig. 2(a)), we perform Nyquist pulse shaping on 2X oversampled symbols using a root-raised cosine (RRC) filter $h_L$ of length $L$ samples. We vary the roll-off factor $\rho$ between 2.5%, 5%, 10%. Then, we digitally pre-compensate for the uneven frequency response of the TX by using a pre-emphasis filter of the same length as the RRC filter (i.e., of length $L$ samples). As is typical, when TX components have a decreasing gain towards higher frequencies, there is a trade-off between recovered SNR and bandwidth utilization; namely, assuming Nyquist pulse shaping, a small symbol rate $R_{Sym}$ can produce a high recovered SNR but can waste available bandwidth, whereas a large $R_{Sym}$ can aggressively utilize the bandwidth but can reduce the recovered SNR. Therefore, we modulate symbols at increasing rates of $R_{Sym}$ = 84.375, 90, 95.625, 101.25, 106.875 GBd and choose the optimal $R_{Sym}$ that yields the greatest achievable NDR. Here, for each $R_{Sym}$, we use the optimal signal attenuation for the pre-emphasis filter, as described in Sec. 3.2. We perform RX DSP in the order of the front-end correction (timing skew compensation and automatic gain control), resampling to 2 samples per symbol, timing recovery [12], frequency-domain PDM equalization using a least-mean square (LMS) filter, frequency offset compensation (FOC), and phase recovery (CPR) based on 2% pilots and blind phase search (BPS) [13].

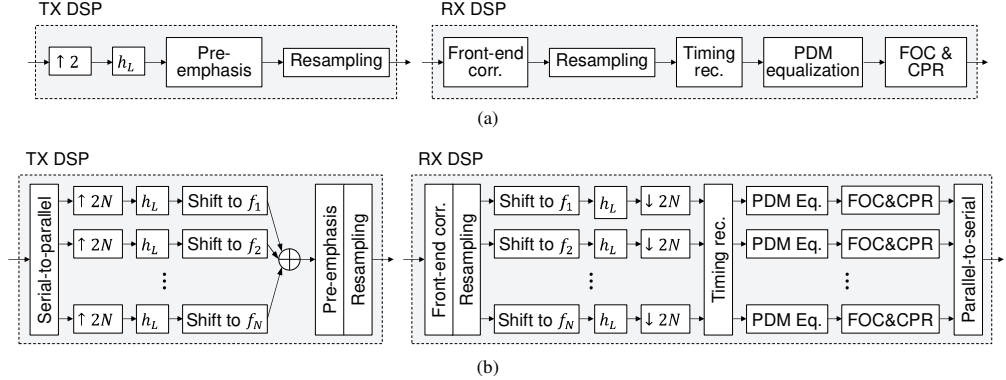

Fig. 2. TX and RX DSP for (a) SC and (b) DSM systems.

For the case of DSM with $N$ subcarriers (Fig. 2(b)), RRC pulse shaping is performed over $2N$-times upsampled signal on a per-subcarrier basis, using the same set of $\rho$ as in SC (i.e., 2.5%, 5%, 10%). Throughout the paper, we use the number of subcarriers $N = 8$ for DSM. For the same set of symbol rates $R_{Sym}$ as in SC, each subcarrier is digitally modulated at $R_{Sym}/N$ GBd. We use the subcarrier spacing of $R_{Sym}/N \times (1 + \rho)$ GHz to avoid inter-subcarrier interference caused by the RRC filter's sidelobes. The subcarriers are shifted to their respective intermediate frequencies $f_n$ for $n = 1, ..., N$, and are combined as shown in Fig. 2(b). In the RX DSP, the signals are resampled at a rate of $2R_{Sym}$ GSa/s and are shifted in frequency to the baseband. Matched filtering is performed on a per-subcarrier basis using the same RRC filters as in the TX. Each subcarrier is downsampled by a factor of $2N$, thereby realizing 2X oversampling on each subcarrier. Timing recovery is performed by averaging the timing errors over two center subcarriers (i.e., the 4th and 5th subcarriers), and the PDM equalization, FOC, and CPR are performed as in SC but on a per-subcarrier basis.

## 3. Analysis of TX DSP for SC and DSM systems

### 3.1 RRC filter

Nyquist pulse shaping plays an important role in band-limited systems, and it is even more so for DSM systems, as we will see in this section. Fig. 3 shows the power spectral density (PSD) of RRC-filtered signals with $\rho = 10\%$ and 2.5% using a large filter length $L$ (such that distortion due to finite filter length is negligible) at $R_{Sym} = 101.25$ GBd. The black dashed lines represent the minimum Nyquist bandwidth having boundaries at $\pm R_{Sym}/2$ GHz for $R_{Sym}$ GBd, and the red solid lines represent the bandwidth to contain 99% of the total signal power. By comparing the black and red lines, we can evaluate how much excess bandwidth is required to carry 99% of the signal power compared to the theoretical minimum of $R_{Sym}$ GHz. As shown in Fig. 3(a), if we use an RRC roll-off factor of $\rho = 10\%$, SC requires only 2 GHz excess bandwidth, but DSM requires much larger 8.8 GHz excess bandwidth. In order to make the DSM occupy a similar excess bandwidth as the SC with $\rho = 10\%$, the RRC roll-off factor $\rho$ of DSM should significantly be reduced to 2.5%, as shown in Fig. 3(b). This shows that unlike the conventional understanding that DSM uses the same bandwidth as SC, *DSM requires more bandwidth than SC* for the same Nyquist pulse shaping.

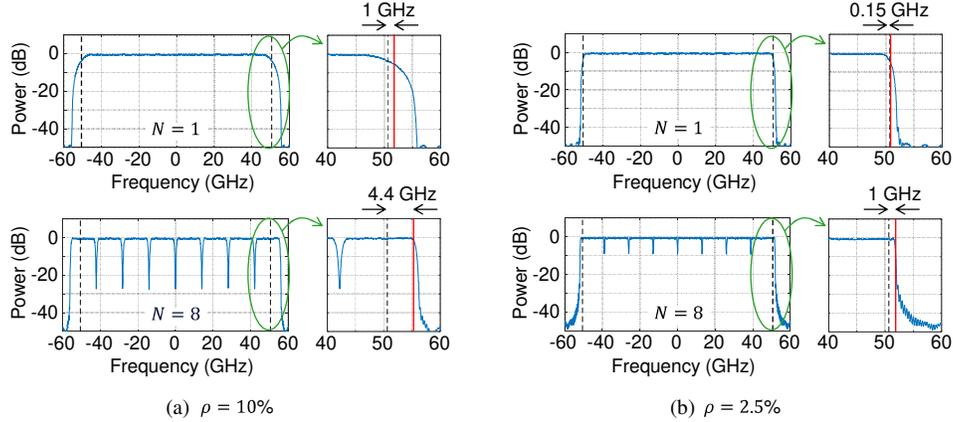

Fig. 3. PSD of RRC filtered signals with (a) $\rho = 10\%$ and (b) $\rho = 2.5\%$, in SC (top) and DSM (bottom) systems. The black vertical dashed lines indicate $\pm R_{Sym}/2$ GHz, and the red vertical solid lines indicate the bandwidth to contain 99% of the total signal energy.

Using a small $\rho$ can alleviate the bandwidth expansion problem of DSM. Note, however, that the filter length $L$ should increase as $\rho$ decreases *and $N$ increases*, to avoid increased distortion due to RRC filtering. More specifically, to maintain a similar level of distortion, $L$ should increase almost linearly with $1/\rho$ and exactly linearly with $N$. For example, DSM with $\rho = 2.5\%$ and $N = 8$ requires ~32X larger $L$ than SC with $\rho = 10\%$. The reason that $L$ increases linearly with $N$ is that RRC filtering should be performed on the $2N$ times upsampled data, as shown in Fig. 2(b), so the number of symbols over which the filter spans decreases to $L/(2N)$ as $N$ increases.

Fig. 4(a) shows a straight-forward implementation of the length-$L$ RRC filter $H(Z) = \sum_{k=0}^{L-1} h_L(k) Z^{-k}$, where the numbers in the parentheses represent the clock rate for digital circuitry. It can easily be seen that RRC filtering (in the blue box) requires $2N$ complex multiplications per subcarrier per symbol time per polarization. Since there are $N$ identical RRC filters in the TX for DSM, $2LN$ complex multiplications are required per symbol time per polarization, which increases linearly with $N$. Fig. 4(b) illustrates an efficient polyphase implementation [14] of Fig. 4(a), where the $n$-th of the $2N$ blue boxes performs length-$(L/2N)$ RRC filtering with $H_n(Z) = \sum_{k=0}^{L/(2N)-1} h_{L/(2N)}^{(n)}(k) Z^{-2Nk}$ and $h_{L/(2N)}^{(n)}(k) \triangleq h_L(2Nk + n)$, and a subsequent parallel-to-serial converter is responsible for applying a proper delay $Z^{-n}$ for each $n$. For a given $\rho$, the polyphase implementation effectively reduces the complexity of the RRC filtering from $2LN$ to $L$ complex multiplications per symbol per polarization, removing the dependency on $N$. Nevertheless, as mentioned above, a smaller $\rho$ is needed for the DSM to prevent the bandwidth expansion problem, and this increases the complexity of RRC filtering as described below.

Fig. 5 shows the frequency response of the RRC filters for $R_{Sym} = 101.25$ GBd, with $\rho = 2.5\%$ (green lines), 5% (orange lines), and 10% (blue lines). The top figures show the frequency response at 0 to 100 GHz in the baseband, the middle figures show the enlarged passband response, and the bottom figures show the enlarged transition and stop bands response. For the case of SC (Fig. 5(a)), the shortest RRC filter length of $L = 129$ performs well for all $\rho = 2.5\%$, 5%, 10%. However, for the case of DSM, the same filter length $L = 129$ (Fig. 5(b)) produces large ripples in the passband, cannot realize the targeted cutoff frequency at the transition band (especially for $\rho = 2.5\%$, compare the green lines between Figs. 5(b) and 5(c) in the transition band), and cannot sufficiently suppress the stopband signals (cf. the stopband attenuation is only ~20 dB, see the bottom figure). DSM requires a much larger $L$ to realize the targeted frequency response, as shown in Fig. 5(c) with $L = 513$ for example.

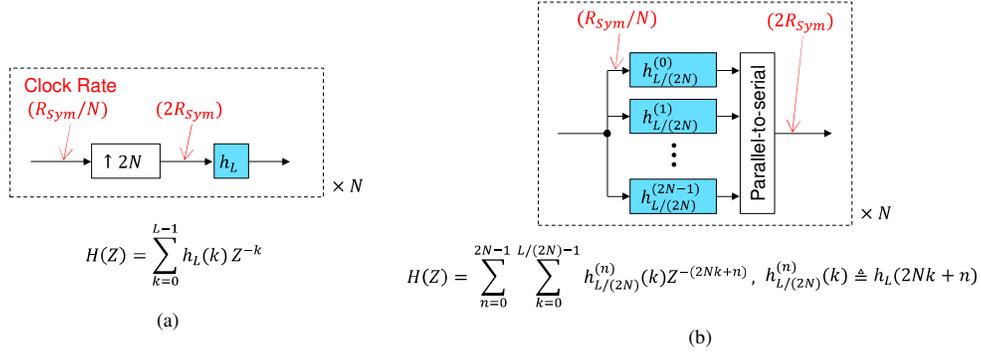

Fig. 4. Implementations of the length-$L$ RRC filter for DSM. (a) A straight-forward implementation, and (b) an efficient polyphase implementation.

To summarize: *(i)* whether it is in an SC or DSM system (i.e., regardless of $N$), a length-$L$ RRC filter for the same $\rho$ can be implemented with the same complexity if a polyphase structure is used, but *(ii)* for an RRC filter of the same $\rho$ to achieve similar performance, it requires $N$ times larger filter length $L$ in the DSM system than in the SC system, and *(iii)* to avoid the bandwidth expansion problem, the RRC filter should use a much smaller $\rho$ in the DSM system than in the SC system, thus further increasing the complexity.

*3.2 Pre-emphasis filter*

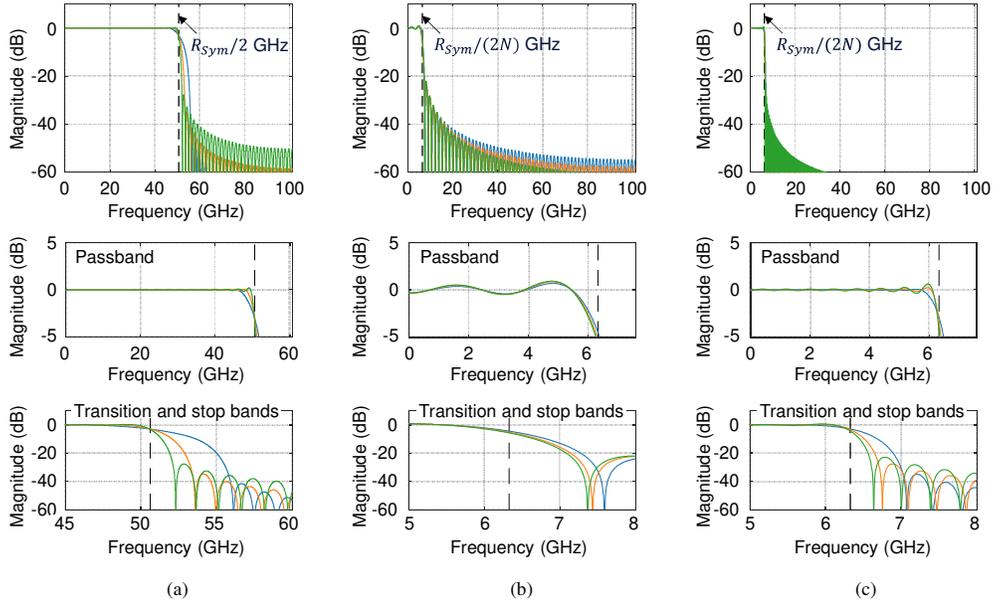

Fig. 5. Frequency response of RRC filters: (a) SC, $L = 129$, (b) DSM, $L = 129$, (c) DSM, $L = 513$, with $\rho = 2.5\%$ (green), 5% (orange), 10% (blue).

In addition to the RRC filters, digital pre-emphasis filters [15-17] are widely used in band-limited optical systems to compensate for the uneven frequency responses of TX modules, since DACs and MZMs typically have a gradually decreasing gain towards higher frequencies. A digital pre-emphasis filter attenuates the low-frequency components of digital signals, thus *relatively* enhancing the high-frequency components that will not receive sufficient gain from subsequent TX modules. The pre-emphasis filter allows us to aggressively increase $R_{Sym}$ and utilize a wider bandwidth for the same TX modules, thereby producing a higher NDR per optical carrier than without a pre-emphasis filter. However, the use of a pre-emphasis filter introduces complications to the NDR maximization problem. The optimal pre-emphasis filter shape depends on the bandwidth occupancy of the signal (i.e., it depends on all of $R_{Sym}$, $\rho$, and $N$), and the total power and power spectral density of TX noise can change with the pre-emphasis filter shape. Therefore, in the presence of the pre-emphasis filter, the linearity of the channel does not hold, invalidating the assumption for water filling and entropy loading to be optimal. In this work, we first create a pre-emphasis filter that has the inverse magnitude response of TX, and limit the maximum attenuation $G_{clip}$ (at the zero frequency) to 10, 12, …, 20 dB to find best compromise between the frequency flatness and TX noise suppression. Then, we choose the optimal $G_{clip}$ that leads to the greatest NDR for each $R_{Sym}$, $\rho$, and $N$. The pre-emphasis filter is implemented in the time domain, and the filter length is chosen to be the same as that of the RRC filter. This allows to merge the RRC and pre-emphasis filters into a single length-$L$ filter.

Fig. 6(a) shows the PSDs (top and middle figures) and recovered SNRs (bottom figures) of the pre-emphasis filtered signals with $G_{clip} = 10$ (left figures) and 20 dB (right figures), after RRC filtering with $\rho = 2.5\%$. The measurements are obtained in the B2B setup of Fig. 1 without ASE noise loading. In the left-hand side figures of Fig. 6(a), the maximum attenuation $G_{clip}$ at the center frequency is relatively small (top), and thus the RX signal is measured with weak power near the band edges (middle). Correspondingly, the recovered SNRs are much lower in edge subcarriers than in center subcarriers (bottom). On the other hand, as shown in the right-hand side figures of Fig. 6(a), if $G_{clip}$ at the center frequency is large (top), the RX signal has increased power near the band edges and hence more closely realizes the water-filling principle (middle). However, a large $G_{clip}$ reduces the average signal power at the DAC output at the expense of flattening the frequency response, and therefore it increases the TX noise power created by the modulator, driver, and EDFA following the DAC [16]. It also increases the quantization noise in the low-frequency regime [15,17] which can affect the performance of high-order QAM. As a result, compared with a smaller $G_{clip}$, a large $G_{clip}$ leads to smaller recovered SNRs at the center subcarriers at the cost of larger recovered SNRs at the edge subcarriers (bottom). This exemplarily shows that the TX noise is not merely additive but varies depending on the pre-emphasis filter shape, negating the optimality of the water-filling [10,11] in the presence of the pre-emphasis filter.

Fig. 6(b) shows the impact of $G_{clip}$ on the achievable NDR with two different bandwidth utilizations of $R_{Sym} = 84.375$ GBd (upper figure) and 101.25 GBd (lower figure). Denoting the recovered SNR at the $n$-th subcarrier as $SNR_n$, the achievable NDR is approximately determined as $2R_{Sym}/N \cdot \sum_{n=1}^{N} \log_2(1 + SNR_n)$ Gb/s, with $N = 1$ for SC and 8 for DSM. The NDRs thus obtained are accurate at low OSNRs (to be presented in Section 4) where the transmitted PS 16-QAM closely represents the rate-matched modulation format. The NDRs may be slightly overestimated at high OSNRs, but the extent of overestimation would only be marginal. If $R_{Sym}$ is small and there is little bandwidth limitation (upper figure), little impact is observed from optimizing the pre-emphasis filter for all cases, in both SC (solid lines) and DSM (dashed lines) systems. However, as $R_{Sym}$ increases and the system becomes more severely band-limited (lower figure), the NDR can significantly be increased by optimizing the pre-emphasis filter, and the optimal filter shape is determined by a complex interplay between $\rho$ and $G_{clip}$, as detailed in the next section. This indicates that it is not immediately clear in realistic band-limited optical systems whether DSM achieves a larger NDRs than SCs, and thus we perform exhaustive search over $R_{Sym}$, $\rho$, and $G_{clip}$ to maximize the NDR in this work. Note that in Fig. 6(b), DSM yields smaller NDRs than SC, and little effect of $G_{clip}$ on NDR is observed in DSM. This is because in the DSM system, the signal distortion from length-129 RRC filtering plays a bigger role than the bandwidth limitation, unlike in the SC system where distortion from length-129 RRC filtering is negligible.

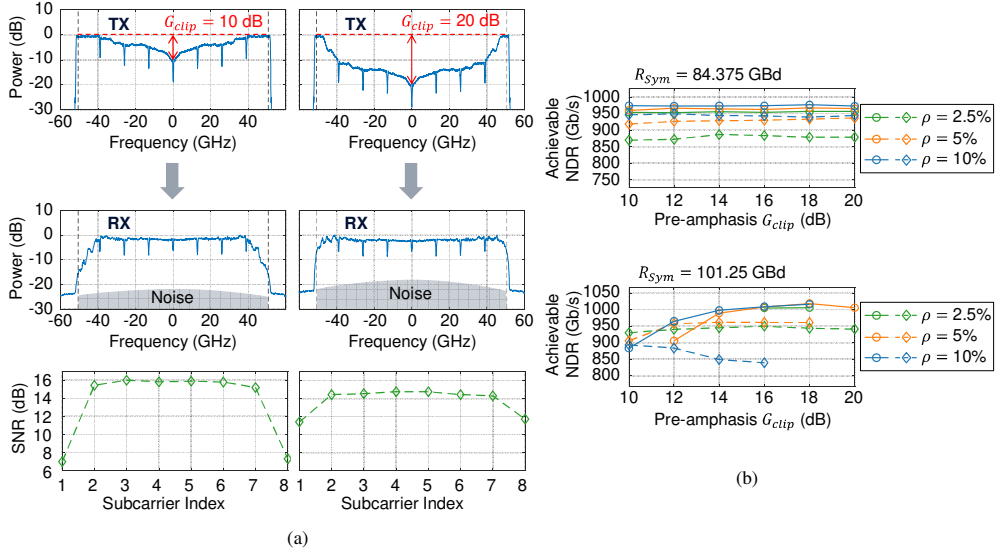

Fig. 6. (a) PSD of the pre-emphasis filtered signals (top), PSD of the corresponding RX signals (middle), and the recovered SNR at each subcarrier (bottom), with $G_{clip} = 10$ dB (left) and 20 dB (right) at $R_{Sym} = 101.25$ GBd. (b) Achievable NDR as a function of $G_{clip}$ at $R_{Sym} = 84.375$ GBd (upper) and 101.25 GBd (lower), with SC (solid lines) and DSM (dashed lines). $L = 129$ for RRC filtering.

### 4. Experimental results

We evaluate the performance of the SC and DSM systems using the experimental setup of Fig. 1, at OSNRs of 8.9 dB, 14.9 dB, 20.9 dB, and 24.9 dB. All OSNRs in this work are measured with the same noise bandwidth as $R_{Sym}$ GHz, which represent the theoretical limit of recovered SNR. Fig. 1(b) shows the typical recovered signal constellations at high OSNR.

Fig. 7 shows the recovered SNR at $R_{Sym} = 84.375$ GBd (top figures) and $R_{Sym} = 95.625$ GBd (bottom figures) in the SC (solid lines) and DSM (dashed lines) systems, obtained with the NDR-maximizing $G_{clip}$. Denoting $L$ used for the SC and DSM systems by $L_{SC}$ and $L_{DSM}$, respectively, Figs. 7(a) shows the results of $L_{SC} = L_{DSM} = 129$ at OSNR = 24.9 dB. Here, changing $\rho$ produces insignificant differences in SNR in the SC system, indicating that $L_{SC} = 129$ is large enough even for the smallest $\rho = 2.5\%$, as can be expected from Fig. 5(a). On the other hand, in the DSM system, since small $L_{DSM} = 129$ fails to achieve the target cutoff frequency of the RRC filter for small roll-off factors of $\rho = 2.5\%$ and $5\%$, substantial SNR penalty is observed at the center subcarriers. At the same time, for the largest roll off factor of $\rho = 10\%$ that causes the greatest bandwidth expansion, large SNR penalty is observed at the edge subcarriers. Fig. 7(b) shows the results when $L_{DSM}$ is increased to 513 while all the other setup is the same as in Fig. 7(a). Now, DSM no longer suffers excessive

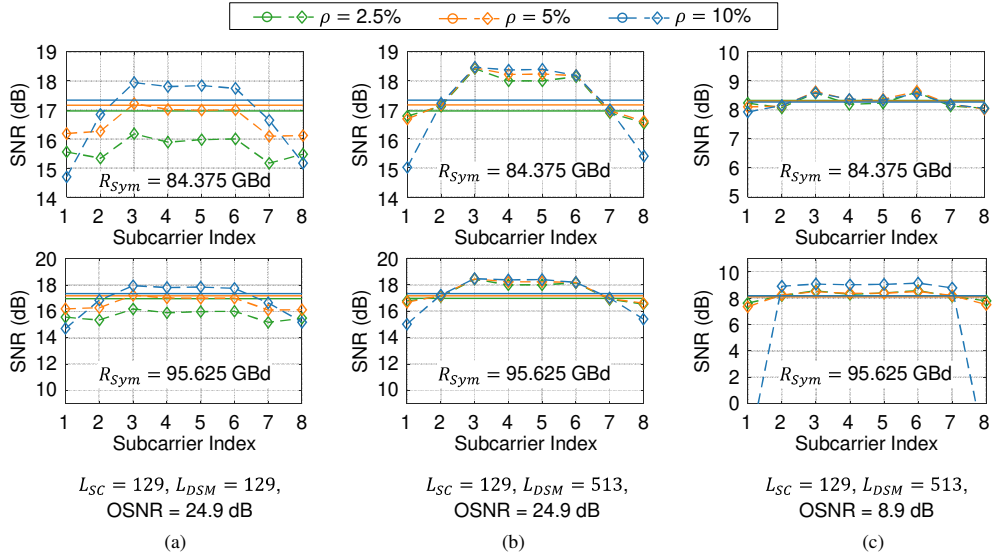

Fig. 7. The recovered SNR of SC (solid lines) and DSM (dashed lines) at $R_{Sym}$ = 84.375 GBd (top) and 95.625 GBd (bottom). $L_{SC}$ = 129 in all cases. (a) $L_{DSM}$ = 129, OSNR = 24.9 dB; (b) $L_{DSM}$ = 513, OSNR = 24.9 dB; (c) $L_{DSM}$ = 513, OSNR = 8.9 dB.

SNR penalty from the signal distortion from RRC filtering. But at the edge subcarriers, the SNR is still the smallest for $\rho = 10\%$ due to the same bandwidth limitation as before. Fig. 7(c) shows the results when OSNR is reduced from 24.9 dB to 8.9 dB while all the other setup is the same as in Fig. 7(b). In this case, the inline ASE noise is much larger than the TX noise and predominantly determines the recovered SNR. Therefore, the SC and DSM systems achieve similar SNRs when $R_{Sym}$ = 84.375 GBd, where the bandwidth limitation is relaxed. When $R_{Sym}$ = 95.625 GBd and the bandwidth limitation is severer, almost the same results are observed as for $R_{Sym}$ = 84.375 GBd, except for $\rho = 10\%$. It should be noted that $\rho = 10\%$ in the DSM system achieves a greater NDR if it uses a smaller $G_{clip}$ than those of $\rho = 2.5\%$ and 5%. This is because for $\rho = 10\%$, the edge subcarriers are not recovered with high SNR even with a large $G_{clip}$ due to the excessively large bandwidth expansion, and therefore the benefit of improving the SNR on the center subcarriers by reducing $G_{clip}$ (see the SNR of $\rho = 10\%$ is larger than those of $\rho = 2.5\%$ and 5% on the center subcarriers) outweighs the resulting loss due to completely discarding the edge subcarriers. This explains the bottom figure of Fig. 7(c).

Fig. 8 shows the achievable information rate (AIR) estimated as $\sum_{n=1}^{N}\log_2(1+SNR_n)/N$ in b/sym/pol (top figures) and the achievable NDR calculated as $2R_{Sym} \times AIR$ in Gb/s (bottom figures), obtained from the same data as in Fig. 7. With $L_{SC} = L_{DSM} = 129$ (Fig. 7(a)), the SC system produces greater AIR and NDR than the DSM system at high OSNR. When $L_{DSM}$ is increased to 513 (Fig. 7(b)), the DSM system can produce larger AIR and NDR than the SC system at high OSNR, at the cost of the increased TX DSP complexity. At low OSNR corresponding to long-haul transmission (Fig. 7(c)), the DSM system achieves almost the same AIR and NDR as the SC system, even with the increased TX DSP complexity. Note that $\rho = 10\%$ for DSM (blue dashed lines) yields almost the same AIR at 101.25 and 106.875 GBd in Figs. 8(a) and 8(b), and at 95.625 GBd to 106.875 GBd in Fig. 8(c). This is because at these symbol rates, the DSM achieves nearly identical recovered SNRs on the 2nd to 7th subcarriers by giving up utilizing the edge subcarriers. If $R_{Sym}$ is continuously increased to above 106.875 GBd, DSM would also discard the 2nd and 7th subcarriers, which will allow good recovered SNRs at the 3rd to 6th subcarriers to be maintained up to even larger $R_{Sym}$. For the same AIR, NDR increases in proportion to $R_{Sym}$, which leads to the blue dashed lines in the bottom figures.

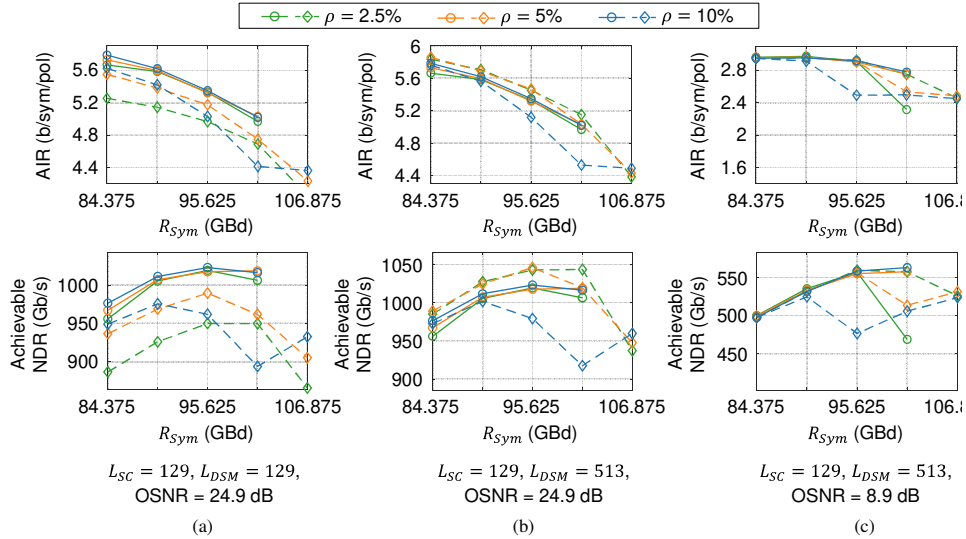

Fig. 8. The AIR (top) and achievable NDR (bottom) of the SC (solid lines) and DSM (dashed lines) systems as a function of $R_{Sym}$. (a) $L_{DSM} = 129$, OSNR = 24.9 dB; (b) $L_{DSM} = 513$, OSNR = 24.9 dB; (c) $L_{DSM} = 513$, OSNR = 8.9 dB. $L_{SC} = 129$ in all cases.

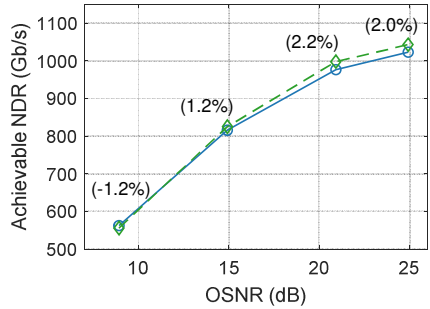

Fig. 9. The achievable NDR of the SC (solid line) and DSM (dashed line) systems as a function of OSNR. The numbers in parentheses indicate the NDR gain of DSM over SC as a percentage.

The achievable NDR maximized over $R_{Sym} \in \{84.375, 90, 95.625, 101.25, 106.875\}$ GBd, $\rho \in \{2.5, 5, 10\}\%$, and $G_{clip} \in \{10, 12, ..., 20\}$ dB is shown in Fig. 9 as a function of the OSNR. As can be seen in the figure, the increase in NDR brought by DSM compared to SC under realistic bandwidth limitations is found to be negligible to marginal for all OSNRs despite the advantage of ideal entropy loading.

## 5. Conclusion

In this work, we have shown that it is a misconception that DSM occupies the same signal bandwidth as SC for the same Nyquist pulse shaping, and that DSM indeed causes a bandwidth expansion problem due to subcarrier spacing. The use of a small roll-off factor $\rho$ can alleviate the bandwidth problem of DSM, but it requires high implementation cost because the RRC filter length must be increased by *MN* times for similar performance, where *M* is the reduction ratio of $\rho$ and *N* is the number of subcarriers. Even if the merits of increased TX complexity and ideal entropy loading are given to DSM, the experimental results have shown that the maximum NDR achievable with DSM is not noticeably larger than that of SC at high OSNR when disregarding the nonlinear fiber propagation effect, and that any marginal NDR gain of DSM vanishes with increasing ASE noise power.

**Disclosures.** The authors declare no conflicts of interest.